\documentclass{article}%
\usepackage{amsfonts}
\usepackage{amsmath}
\usepackage{amssymb}
\usepackage{graphicx}%
\providecommand{\U}[1]{\protect\rule{.1in}{.1in}}
\begin{document}

\title{Quantum coherence in an all-solid-state dye-sentizied solar cell}
\author{C. Benedek\\SolarTech non-profit Ltd. Vadasz str. 5., Szeged, Hungary;\\Department of Theoretical Physics, University of Szeged,\\Tisza Krt. 84., Szeged, Hungary}
\maketitle

\begin{abstract}
The reported new type of all-solid-state, inorganic solar cell will be
discussed by a semiclassical light-matter interaction method. The molecular
compound will be treated by a three times\ two-level coupled quantum system.
The equation of motion of the density matrix of this system will be analytical
solved, in linear approximation and due to the coherent superposition of
certain states, time-independent\ off-diagonal elements will be obtained.
These elements represent an important components for the overal optical
performane of this cell.

\end{abstract}

The dye-sensitized solar cell (DSC) \cite{dsc} is a lowcost and
environmentally friendly alternative for the solid-state devices. It is
inexpensive to produce, and the light-weight thin-film structures are
compatible with automated manufacturing. The new cell improves the
photocurrent density and the power conversion efficiency, this latter is
approx. 10.2\% and it is the first example of an all solid-state
dye-sensitized solar cell that may exceed the performance of a liquid
electrolyte Gr\"{a}tzel cell.

The reported new type of all-solid-state, inorganic solar cell \cite{nature},
that consists of the p-type direct bandgap semiconductor CsSnI$_{3}$ and
n-type nanoporous TiO$_{2}$ with the dye N719
(cisdiisothiocyanato-bis(2,29-bipyridyl-4,49-dicaboxylato) ruthenium(II)
bis-(tetrabutylammonium)),\ will be discussed by a semiclassical light-matter
interaction method.\ In the works \cite{scully}, Scully et. al argue that
quantum coherence can increase the quantum efficiency of various
light-harvesting systems such as for the solar cells.\ This molecular compound
will be treated by a three times\ two-level coupled quantum system. The
equation of motion of the density matrix will be analytical solved, in linear
approximation and time-independent\ off-diagonal elements will be obtained, as
a manifestation of the quantum coherence, that a coherent superposition of
certain states.\ This can be represented an important components for the
overall optical performance of the cell.

A considered two-level quantum system consist of the ground- (HOMO: highest
occupied molecular orbital) and the excited (LUMO: lowest unoccupied molecular
orbital)\ levels.\ The gaps among the considered energy-levels are the next
ones: 3.2 eV for the TiO$_{2}$, 2.37 eV for the dye N719 and 1.3 eV for the
CsSnI$_{3}$. These gaps are allowed by optical transitions. The energy levels
are taken from the literature \cite{el1},\cite{el2}. \ Coupling of the three
two-level quantum system is performed by the dye N719. The dye N719 couples
its LUMO-level with the LUMO-levels of the CsSnI$_{3}$ and TiO$_{2}$ and its
HOMO-level with the HOMO-levels of the CsSnI$_{3}$ and TiO$_{2}$ playing an
"electron-hole transmitter" as in the solid-state organic or p-type conducting
polymer hole-transport materials (HTMs)\ \cite{htm1}, \cite{htm2}. Coupling
coefficients are introduced in order to describe these transitions. Denoting
the molecule TiO$_{2}$ by "T", the dye molecule N719 by "D" and the molecule
CsSnI$_{3}$ by "Cs", the state of the system is:%
\begin{equation}
\left\vert \Psi\right\rangle =C_{g}^{T}\left\vert g\right\rangle _{T}%
+C_{e}^{T}\left\vert e\right\rangle _{T}+C_{g}^{D}\left\vert g\right\rangle
_{N}+C_{e}^{D}\left\vert e\right\rangle _{N}+C_{g}^{Cs}\left\vert
g\right\rangle _{Cs}+C_{e}^{Cs}\left\vert e\right\rangle _{Cs}.\label{S}%
\end{equation}
Where "g" and "e" denote the ground- and excited levels. The density matrix
$\rho=\left\vert \Psi\right\rangle \left\langle \Psi\right\vert $ is a
$6\times6$ matrix. Equation of motion of the density matrix from
\cite{QOScully}:
\begin{equation}
\rho=-\frac{i}{\hbar}\left[  H,\rho\right]  -\frac{1}{2}\left\{  \Gamma
,\rho\right\}
\end{equation}
with the relaxation matrix $\Gamma$, which is a diagonalized matrix and the
Hamiltonian $H$, the i\textit{j}th matrix element of $\rho$%
\begin{equation}
\rho_{ij}=-\frac{i}{\hbar}\sum\left(  H_{ik}\rho_{kj}-\rho_{ik}H_{kj}\right)
-\frac{1}{2}\sum\left(  \Gamma_{ik}\rho_{kj}-\rho_{ik}\Gamma_{kj}\right)
,\label{edm}%
\end{equation}
and%
\begin{equation}
H=H_{0}+H_{I}%
\end{equation}
where%
\begin{align}
H_{0} &  =\hbar\omega_{g}^{T}\left\vert g\right\rangle \left\langle
g\right\vert _{T}+\hbar\omega_{e}^{T}\left\vert e\right\rangle \left\langle
e\right\vert _{T}+\hbar\omega_{g}^{D}\left\vert g\right\rangle \left\langle
g\right\vert _{N}+\nonumber\\
&  \hbar\omega_{e}^{D}\left\vert e\right\rangle \left\langle e\right\vert
_{N}+\hbar\omega_{g}^{Cs}\left\vert g\right\rangle \left\langle g\right\vert
_{Cs}+\hbar\omega_{e}^{Cs}\left\vert e\right\rangle \left\langle e\right\vert
_{Cs}%
\end{align}
\newline%
\begin{align}
H_{I} &  =-\frac{\hbar}{2}(\Omega_{RT}e^{-i\phi_{T}t}e^{-i\omega_{T}%
t}\left\vert g\right\rangle \left\langle e\right\vert _{T}+\Omega
_{RN}e^{-i\phi_{D}t}e^{-i\omega_{D}t}\left\vert g\right\rangle \left\langle
e\right\vert _{D}+\nonumber\\
&  +\Omega_{RCs}e^{-i\phi_{Cs}t}e^{-i\omega_{Cs}t}\left\vert g\right\rangle
\left\langle e\right\vert _{Cs})\\
&  +\alpha_{g}^{TD}\left\vert g\right\rangle _{T}\left\langle g\right\vert
_{D}+\alpha_{e}^{TD}\left\vert e\right\rangle _{T}\left\langle e\right\vert
_{D}+\alpha_{g}^{DCs}\left\vert g\right\rangle _{D}\left\langle g\right\vert
_{Cs}+\alpha_{e}^{DCs}\left\vert e\right\rangle _{D}\left\langle e\right\vert
_{Cs}.\nonumber
\end{align}
Here the $\Omega_{RT}e^{-i\phi_{T}t}$ and $\Omega_{RD}e^{-i\phi_{D}t}\ $and
$\Omega_{RCs}e^{-i\phi_{Cs}t}$ are the complex Rabi frequencies associated
with the coupling of the field modes of frequencies $\omega_{T}$, $\omega_{D}$
and $\omega_{Cs}$ to the atomic transitions $\left\vert g\right\rangle
_{T}\rightarrow\left\vert e\right\rangle _{T},\ \left\vert g\right\rangle
_{D}\rightarrow\left\vert e\right\rangle _{N}\ $and $\left\vert g\right\rangle
_{Cs}\rightarrow\left\vert e\right\rangle _{Cs}\ $respectively.$\ $It has
assumed that only these transitions are directly optical allowed. The
transitions $\left\vert g\right\rangle _{T}\rightarrow\left\vert
g\right\rangle _{D}\ $, $\left\vert e\right\rangle _{T}\ \rightarrow\left\vert
e\right\rangle _{D}$ , $\left\vert g\right\rangle _{D}\rightarrow\left\vert
g\right\rangle _{Cs}\ $and $\left\vert e\right\rangle _{D}\ \rightarrow
\left\vert e\right\rangle _{Cs}$ are not directly optical allowed. The
electron-hole transports among these states are powered by the light absorbing
of the\ dye N719 (the transition $\left\vert g\right\rangle _{D}%
\rightarrow\left\vert e\right\rangle _{D}$ is optical allowed), and\ it
carries electron and holes to the TiO$_{2}$ and CsSnI$_{3}$ molecules
\cite{nature}. Coupling coefficients are introduced, $\alpha_{g}^{TD}%
,\ \alpha_{e}^{TD},$ $\alpha_{g}^{DCs},$ and $\alpha_{e}^{DCs}$\ for the
transitions $\left\vert g\right\rangle _{T}\rightarrow\left\vert
g\right\rangle _{D}$, $\left\vert e\right\rangle _{T}\rightarrow\left\vert
e\right\rangle _{D}$, $\left\vert g\right\rangle _{D}\rightarrow\left\vert
g\right\rangle _{Cs},\ \left\vert e\right\rangle _{D}\rightarrow\left\vert
e\right\rangle _{Cs}.$

The solutions of the off-diagonal density matrix elements in linear
approximation where the molecular compound is initially in the ground level,
are the next: For the matrix element $C_{g}^{T}\left(  C_{e}^{T}\right)
^{\ast}=\rho_{ge}^{T}=\rho_{12}=\widetilde{\rho}_{12}e^{-i\omega_{T}t}%
,$\ denoting with $\omega_{eg}^{T}=\omega_{e}^{T}-$\ $\omega_{g}^{T},$ the
detuning of the field is $\Delta_{T}=\omega_{eg}^{T}-\omega_{T}$\ the decay
rate is denoted by $\gamma_{T}=\left(  \gamma_{g}^{T}+\gamma_{e}^{T}\right)
/2$\ and $d_{T}$\ is the transition dipole moment, field amplitude
$\varepsilon_{T}$ for this transition,%
\begin{equation}
\widetilde{\rho}_{12}=\frac{id_{T}\varepsilon_{T}}{2\hbar\left(  \gamma
_{T}+i\Delta_{T}\right)  }.\label{r12sol}%
\end{equation}
Similarly, for the $\rho_{34}=\widetilde{\rho}_{34}e^{-i\omega_{D}t}=\rho
_{ge}^{D}=C_{g}^{D}\left(  C_{e}^{D}\right)  ^{\ast}$\ matrix element,
denoting $\omega_{eg}^{D}=\omega_{e}^{D}-$\ $\omega_{g}^{D},$\ detuning is
$\Delta_{D}=\omega_{eg}^{D}-\omega_{D},$ the decay rate is denoted by
$\gamma_{D}=\left(  \gamma_{g}^{D}+\gamma_{e}^{D}\right)  /2$\ and $d_{D}$\ is
the transition dipole moment, field amplitude $\varepsilon_{D}$ for this
transition,%
\begin{equation}
\widetilde{\rho}_{34}=\frac{id_{N}\varepsilon_{N}}{2\hbar\left(  \gamma
_{N}+i\Delta_{N}\right)  }.\label{r34sol}%
\end{equation}
And\ for the $\rho_{56}=\widetilde{\rho}_{56}e^{-i\omega_{Cs}t}=\rho_{ge}%
^{Cs}=C_{g}^{Cs}\left(  C_{e}^{Cs}\right)  ^{\ast}$\ matrix element, denoting
$\omega_{eg}^{Cs}=\omega_{e}^{Cs}-$\ $\omega_{g}^{Cs},$ detuning is
$\Delta_{Cs}=\omega_{eg}^{Cs}-\omega_{Cs},$ the decay rate is denoted by
$\gamma_{Cs}=\left(  \gamma_{g}^{Cs}+\gamma_{e}^{Cs}\right)  /2$ and $d_{Cs}%
$\ is the transition dipole moment, field amplitude $\varepsilon_{Cs}$ for
this transition,\newline%
\begin{equation}
\widetilde{\rho}_{56}=\frac{id_{Cs}\varepsilon_{Cs}}{2\hbar\left(  \gamma
_{Cs}+i\Delta_{Cs}\right)  }.\label{r56sol}%
\end{equation}

Equation of motion for the $\rho_{24}=\rho_{ee}^{TD}=C_{e}^{T}\left(
C_{e}^{D}\right)  ^{\ast}$ matrix element, denoting $\omega_{ee}^{TD}%
=\omega_{e}^{T}-$\ $\omega_{e}^{D},$the decay rate is denoted by $\gamma
_{TD}^{e}=\left(  \gamma_{e}^{T}+\gamma_{e}^{D}\right)  /2$, and solution:
\begin{equation}
\overset{.}{\rho_{24}}=-\left(  \gamma_{TD}^{e}+i\omega_{ee}^{TD}\right)
\rho_{24},
\end{equation}%
\begin{equation}
\rho_{24}=\exp\left[  -\left(  \gamma_{TD}^{e}+i\omega_{ee}^{TD}\right)
t\right]  .\label{r24sol}%
\end{equation}
Equation of motion for the $\rho_{46}=\rho_{ee}^{DCs}=C_{e}^{D}\left(
C_{e}^{Cs}\right)  ^{\ast}$ matrix element, denoting $\omega_{ee}^{DCs}%
=\omega_{e}^{D}-$\ $\omega_{e}^{Cs},$ the decay rate is denoted by
$\gamma_{DCs}^{e}=\left(  \gamma_{e}^{D}+\gamma_{e}^{Cs}\right)  /2$, and
solution:%
\begin{equation}
\overset{.}{\rho_{46}}=-\left(  \gamma_{DCs}^{e}+i\omega_{ee}^{DCs}\right)
\rho_{46},
\end{equation}%
\begin{equation}
\rho_{46}=\exp\left[  -\left(  \gamma_{DCs}^{e}+i\omega_{ee}^{DCs}\right)
t\right]  .\label{r46sol}%
\end{equation}
Equation of motion for the $\rho_{13}=\rho_{gg}^{TD}=C_{g}^{T}\left(
C_{g}^{D}\right)  ^{\ast}$ matrix element, denoting $\omega_{gg}^{TD}%
=\omega_{g}^{T}-$\ $\omega_{g}^{D},$ the decay rate is denoted by $\gamma
_{TD}^{g}=\left(  \gamma_{g}^{T}+\gamma_{g}^{D}\right)  /2$, and solution:%
\begin{equation}
\overset{.}{\rho_{13}}=-\left(  \gamma_{TD}^{g}+i\omega_{gg}^{TD}\right)
\rho_{13}+\frac{i\alpha_{g}^{TD}}{\hbar},
\end{equation}%
\begin{equation}
\rho_{13}=\frac{i\alpha_{g}^{TD}}{\hbar\left(  \gamma_{TD}^{g}+i\omega
_{gg}^{TD}\right)  }.\label{r13sol}%
\end{equation}
Equation of motion for the $\rho_{35}=\rho_{gg}^{DCs}=C_{g}^{D}\left(
C_{g}^{Cs}\right)  ^{\ast}$ matrix element, denoting $\omega_{gg}^{DCs}%
=\omega_{g}^{D}-$\ $\omega_{g}^{Cs},$ the decay rate is denoted by
$\gamma_{DCs}^{g}=\left(  \gamma_{g}^{D}+\gamma_{g}^{Cs}\right)  /2$, and
solution:%
\begin{equation}
\overset{.}{\rho_{35}}=-\left(  \gamma_{DCs}^{g}+i\omega_{gg}^{DCs}\right)
\rho_{35}+\frac{i\alpha_{g}^{DCs}}{\hbar},
\end{equation}%
\begin{equation}
\rho_{35}=\frac{i\alpha_{g}^{DCs}}{\hbar\left(  \gamma_{DCs}^{g}+i\omega
_{gg}^{DCs}\right)  }.\label{r35sol}%
\end{equation}

The well-known solution can be recognized\ for a two-level system in the
expressions (\ref{r12sol}), (\ref{r34sol}) and (\ref{r56sol}), and one can
gains the susceptibility (Lorentz-model) of a two-level system from these
ones. The matrix elements (\ref{r24sol}), (\ref{r46sol}) vanish in time due to
relaxation processes. The most important elements are the expressions
(\ref{r13sol}) and (\ref{r35sol}), as a coherent superpositions of the ground
levels. These time-independent, not vanishing components represent the
coupling of the three two-level quantum systems and have an influence for the
overall optical performance of the cell.

\end{document}